\documentclass{PoS}

\def\be{\begin{equation}}
\def\ee{\end{equation}}

\newcommand{\ud}{\mathrm{d}}

\title{Statistical hadronization of charm quarks in ultra-relativistic
nucleus-nucleus collisions}

\ShortTitle{Statistical hadronization of charm quarks in ultra-relativistic
nucleus-nucleus collisions}

\author{\speaker{Anton Andronic}%
         \thanks{Also at NIPNE Bucharest.}\\
        GSI, Darmstadt\\
        E-mail: \email{A.Andronic@gsi.de}}

\author{Peter Braun-Munzinger\\
        GSI, Darmstadt and Technical University, Darmstadt\\
        E-mail: \email{P.Braun-Munzinger@gsi.de}}

\author{Krzysztof Redlich\\
        Institute of Theoretical Physics, University of Wroc\l aw and GSI, Darmstadt\\
        E-mail: \email{redlich@ift.uni.wroc.pl}}

\author{Johanna Stachel\\
        Physikalisches Institut der Universit\"at Heidelberg\\
        E-mail: \email{stachel@physi.uni-heidelberg.de}}

\abstract{
We discuss the production of charmonium and open charm in nuclear
  collisions at SPS/FAIR energies within the framework of the statistical
  hadronization model. The increasing importance at lower energies of
  $\Lambda_c$ production is discussed and provides a challenge for future
  experiments. We also show that possible modifications of charmed hadrons 
  in the hot hadronic medium do not lead to measurable changes in the
  cross sections for D-meson production. A possible influence of medium 
  effects can be seen, however, in yields of
  charmonium. These effects are visible at all energies
  and  results will be presented for the energy range between charm threshold
  and RHIC energy. 
}

\FullConference{Critical Point and Onset of Deconfinement   -  4th
International Workshop\\
                 July 9 - 13, 2007\\
                 Darmstadt, Germany}

\begin{document}

\section{Introduction}

Charmonium production is considered, since the original proposal more than 20
years ago about its suppression in a Quark-Gluon Plasma (QGP)
\cite{satz}, as an important probe to determine the degree of deconfinement
reached in the fireball produced in ultra-relativistic nucleus-nucleus
collisions. 
In the original scenario of J/$\psi$ suppression via Debye screening
\cite{satz} it is assumed that the charmonia are rapidly formed in initial
hard collisions but are subsequently destroyed in the QGP (see an update
of this picture in ref. \cite{satz2}).

In a recent series of publications \cite{aa1,aa2,aa3} we have
demonstrated that, in the energy range from top SPS energy ($\sqrt{s_{NN}}
\approx 17$ GeV) on,  the data on J/$\psi$ and $\psi'$ production in
nucleus-nucleus collisions can be well described within the statistical
hadronization model proposed in \cite{pbm1}. 
This includes the centrality and rapidity dependence of recent data at RHIC 
($\sqrt{s_{NN}}$=200 GeV) published by the PHENIX collaboration \cite{phe1}.
We note that the extrapolation of these results to LHC energy 
($\sqrt{s_{NN}}$=5.5 TeV) yields a rather striking centrality dependence 
\cite{aa2,aa3}. Depending on the magnitude of the $\bar c c$ cross section
in central Pb-Pb collisions \cite{aa5}, even an enhancement of 
J/$\psi$ production compared to pp collisions ($R_{AA}^{J/\psi} > 1$) is
expected due to hadronization (at chemical freeze-out) of uncorrelated 
(at these high energies) charm quarks thermalized in QGP.

Here we explore the lower energy range (from near threshold, 
$\sqrt{s_{NN}} \approx 6$ GeV), which can be investigated in 
the CBM experiment \cite{cbm1} at the future FAIR facility. 
One of the motivations for such studies was the expectation \cite{cbm1,tol} 
to provide, by a measurement of D-meson production near threshold, 
information on their possible in-medium modification near the phase boundary. 
However, the cross section $\sigma_{c \bar c}$ is governed 
by the mass of the charm quark $m_c \approx 1.3$ GeV, which 
is much larger than any soft Quantum Chromodynamics (QCD) scale such as 
$\Lambda_{QCD}$. 
Therefore we expect no medium effects on this quantity.\footnote{Such 
a separation of scales is not possible for strangeness production, and
the situation there is not easily comparable.}
The much later formed D-mesons, or other charmed hadrons, may well change
their mass in the hot medium. 
The results of various studies on in-medium modification of charmed hadrons 
masses \cite{tol,tsu,sib1,sib,hay,cas,fri,lutz,mor} are sometimes contradictory.
Whatever the medium effects may be, they can, because of the charm 
conservation, $\sigma_{c \bar c} = \frac{1}{2} ( \sigma_D +
\sigma_{\Lambda_c} +\sigma_{\Xi_c} + ...) + ( \sigma_{\eta_c} +
\sigma_{J/\psi} + \sigma_{\chi_c} + ...)$, only lead to a redistribution 
of charm quarks \cite{aa4}.
This argument is essentially model-independent and applies equally well 
at all energies.  Here we will consider various types of scenarios for 
medium modifications and study their effect within the statistical 
hadronization framework in the energy range from charm threshold to 
collider energies. 
In this context, we note that excellent fits of the common (non-charmed) 
hadrons to predictions of the thermal model have been obtained using vacuum 
masses (see ref. \cite{aat} and references therein). An attempt to use 
modified masses for the RHIC energy \cite{bro} has not produced a conclusive 
preference for any mass or width modifications of hadrons in medium. 
On the other hand, some evidence for possible mass modifications was 
presented in the chiral model of \cite{zschiesche}.
   

\section{Assumptions and ingredients of the statistical
  hadronization model} 

The statistical hadronization model (SHM) \cite{pbm1,aa2} assumes that 
the charm quarks are produced in primary hard collisions and
that their total number stays constant until hadronization.  
Another important factor is thermal equilibration in the QGP, at least 
near the critical temperature, $T_c$. 
We neglect charmonium production in the nuclear corona \cite{aa2},  
since we focus in the following on central collisions ($N_{part}$=350),
where such effects are small. 

In the following we briefly outline the calculation steps in our model
\cite{pbm1,aa2}.  
The model has the following input parameters:
i) charm production cross section in pp collisions;
ii) characteristics at chemical freeze-out: temperature, $T$, 
baryochemical potential, $\mu_b$, and volume corresponding to one unit 
of rapidity $V_{\Delta y=1}$ (our calculations are for midrapidity). 
Since, in the end, our main results will be ratios of hadrons with charm 
quarks nomalized to the $\bar c c$ yield, the detailed magnitude of the 
open charm cross section and whether to use integrated yield or midrapidity 
yields is not crucial.

\begin{figure}[ht]
\vspace{-.7cm}
\begin{tabular}{lc} \begin{minipage}{.62\textwidth}
\hspace{-.4cm}\includegraphics[width=1.2\textwidth,height=1.13\textwidth]{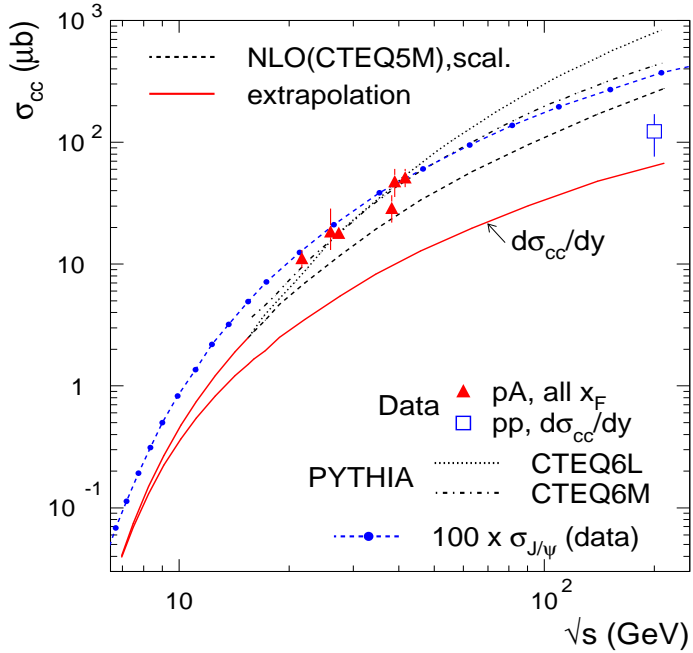}
\end{minipage}  & \begin{minipage}{.34\textwidth}
\caption{Energy dependence of the charm production cross 
section in pp collisions. The NLO pQCD values \cite{rv1} are 
compared to calculations using PYTHIA and to data in pA collisions, taken 
from ref. \cite{lou}. 
Our extrapolations for low energies are shown with continuous lines, 
for total and midrapidity ($\ud\sigma_{c\bar{c}}/\ud y$) cross section.
The open square is a midrapidity measurement in pp collisions \cite{phe3}.
The dashed line with dots indicates a parameterization of the measured energy
dependence of the $J/\psi$ production cross section \cite{herab}.} 
\label{aa_fig0}
\end{minipage} \end{tabular} 
\end{figure}

The charm balance equation \cite{pbm1}, which has to include canonical 
suppression factors \cite{gor} whenever the number of charm pairs is 
not much larger than 1, is used to determine a fugacity  factor $g_c$ via:
\begin{equation}
N_{c\bar{c}}^{dir}=\frac{1}{2}g_c N_{oc}^{th}
\frac{I_1(g_cN_{oc}^{th})}{I_0(g_cN_{oc}^{th})} + g_c^2N_{c\bar c}^{th}.
\label{aa:eq1}
\end{equation}
Here $N_{c\bar{c}}^{dir}$  is the number of initially produced $c\bar{c}$ 
pairs and  $I_n$ are  modified Bessel functions. In the fireball of volume 
$V$ the total number of open ($N_{oc}^{th}=n_{oc}^{th}V$) and hidden 
($N_{c\bar c}^{th}=n_{c\bar c}^{th}V$) charm hadrons is computed from 
their grand-canonical densities $n_{oc}^{th}$ and $n_{c\bar c}^{th}$, 
respectively. This charm balance equation is the implementation within our
model of the charm conservation constraint.
The densities of different particle species in the grand canonical ensemble 
are calculated following the statistical model \cite{aat}.
The balance equation (\ref{aa:eq1}) defines the fugacity parameter $g_c$ 
that accounts for deviations of heavy quark multiplicity from the value 
that is expected in complete chemical equilibrium. 
The yield of charmonia of type $j$ is obtained as: $N_j=g_c^2 N_j^{th}$,
while the yield of open charm hadrons is: 
$N_i=g_c N_i^{th}{I_1(g_cN_{oc}^{th})}/{I_0(g_cN_{oc}^{th})}$.

\begin{figure}[htb]
\begin{tabular}{lr}\begin{minipage}{.49\textwidth}
\hspace{-0.5cm}\includegraphics[width=1.05\textwidth]{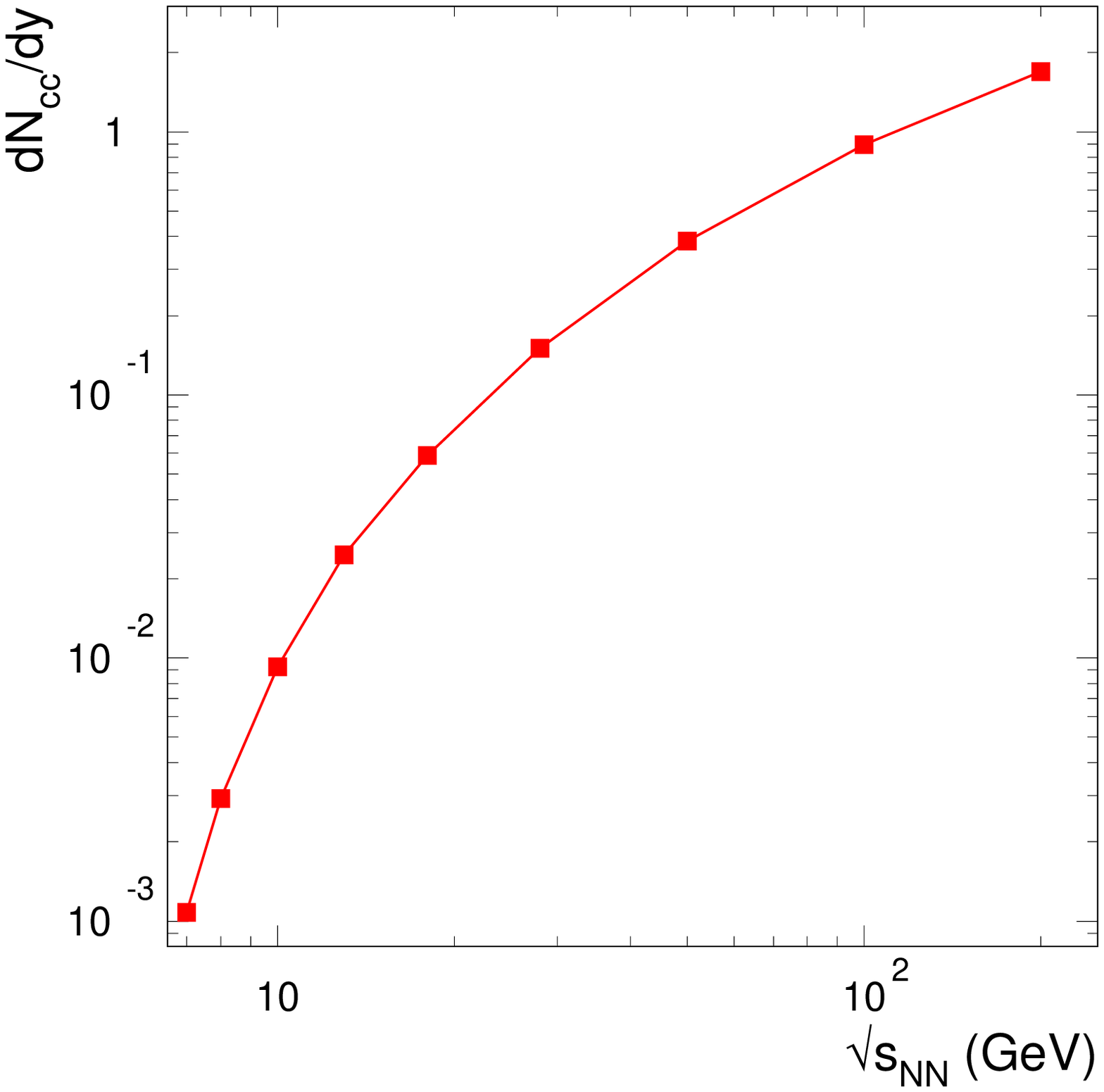}
\caption{Energy dependence of the number of initially produced charm 
quark pairs ($N_{part}$=350).} 
\label{aa_fig1a}
\end{minipage}  & \begin{minipage}{.49\textwidth}
\hspace{-0.5cm}\includegraphics[width=1.05\textwidth]{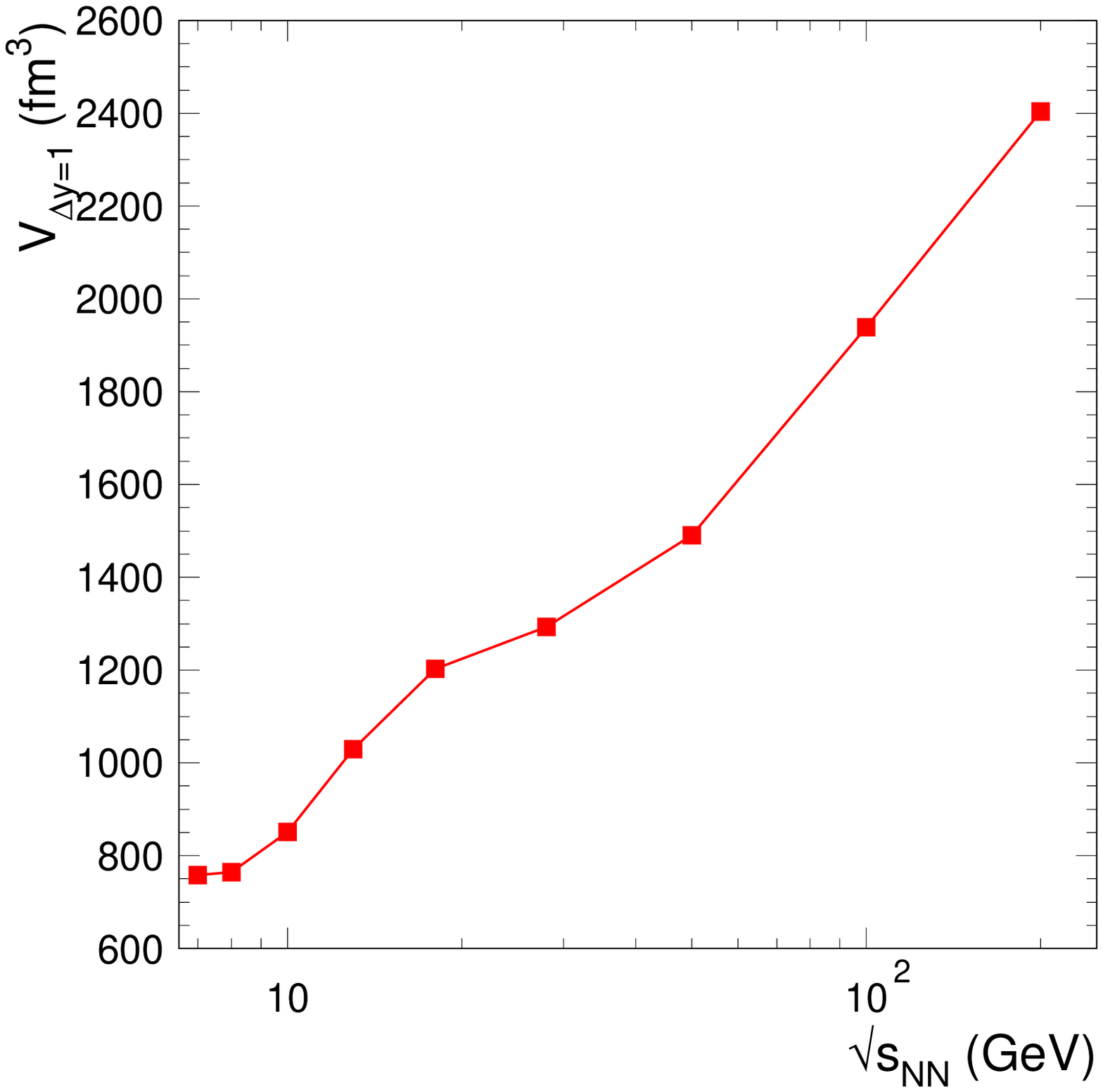}
\caption{Energy dependence of the volume at midrapidity, $V_{\Delta y=1}$,
for central collisions \cite{aat}.} 
\label{aa_fig1b}
\end{minipage} \end{tabular}
\end{figure} 

As no information on the charm production cross section is available for 
energies below $\sqrt{s}$=15 GeV, we have to rely on extrapolation. 
The basis for this extrapolation is the energy dependence of the total 
charm production cross section calculated in ref. \cite{rv1} for the 
CTEQ5M parton distribution functions in next-to-leading order (NLO), 
as shown in Fig.~\ref{aa_fig0}.
We have scaled these calculations to match the more recent values
calculated at $\sqrt{s}$=200 GeV in ref. \cite{cac}.
We employ a threshold-based extrapolation using the following expression:
\begin{equation}
\sigma_{c\bar{c}}=k (1-\sqrt{s_{thr}}/\sqrt{s})^a(\sqrt{s_{thr}}/\sqrt{s})^b
\end{equation}
with $k$=1.85 $\mu$b, $\sqrt{s_{thr}}$=4.5 GeV (calculated assuming a charm
quark mass $m_c$=1.3 GeV \cite{pdg}), $a$=4.3, and $b$=-1.44. 
The parameters $a$, $b$, $k$ were tuned to reproduce the low-energy part 
of the (scaled) NLO curve.
The extrapolated curves for charm production cross section are shown with
continuous lines in Fig.~\ref{aa_fig0}.  Also shown for comparison are
calculations with PYTHIA \cite{lou}.  To obtain the values at midrapidity we
have extrapolated to lower energies the rapidity widths (FWHM) of the charm
cross section known to be about 4 units at RHIC \cite{cac} and about 2 units
at SPS \cite{pbm2}.  With these cross section values, the rapidity density of
initially produced charm quark pairs, shown in Fig.~\ref{aa_fig1a} strongly 
rises from 1.1$\cdot$10$^{-3}$ to 1.7 for the energy range 
$\sqrt{s_{NN}}$=7-200 GeV.  We note that the
so-obtained charm production cross section has an energy dependence similar to
that measured for $J/\psi$ production, recently compiled and parametrized by
the HERA-B collaboration \cite{herab}.  For comparison, this is also shown in
Fig.~\ref{aa_fig0}.  The extrapolation procedure for the low-energy part of
the cross section obviously implies significant uncertainties. We emphasize,
however, that the most robust predictions of our model, i.e. the yields of
charmed hadrons and charmonia relative to the initially produced $c \bar c$
pair yield are not influenced by the details of this extrapolation.

\begin{figure}[hbt]
\begin{tabular}{lr}\begin{minipage}{.49\textwidth}
\hspace{-0.5cm}\includegraphics[width=1.05\textwidth]{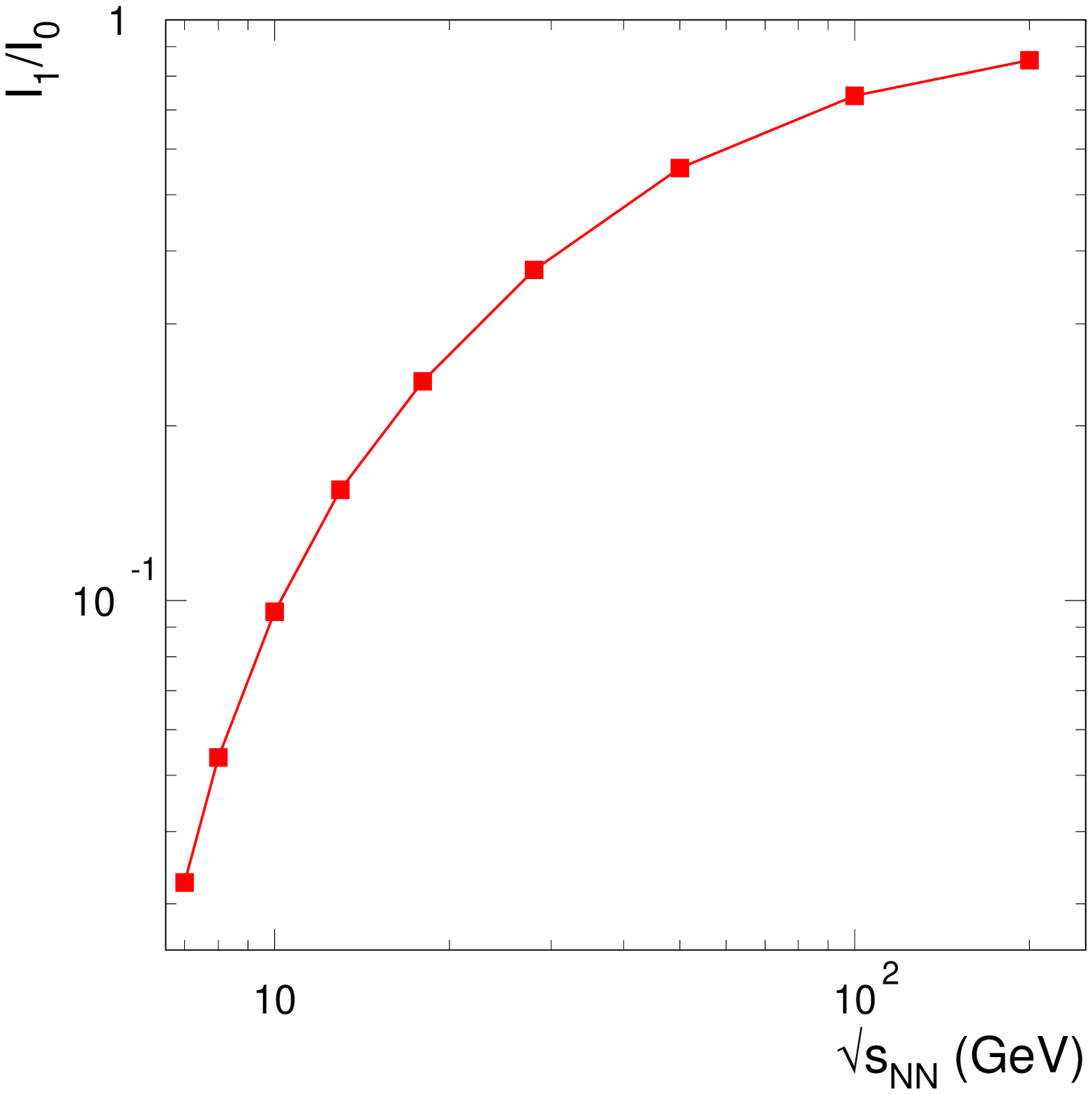}
\end{minipage}  & \begin{minipage}{.49\textwidth}
\hspace{-0.5cm}\includegraphics[width=1.05\textwidth]{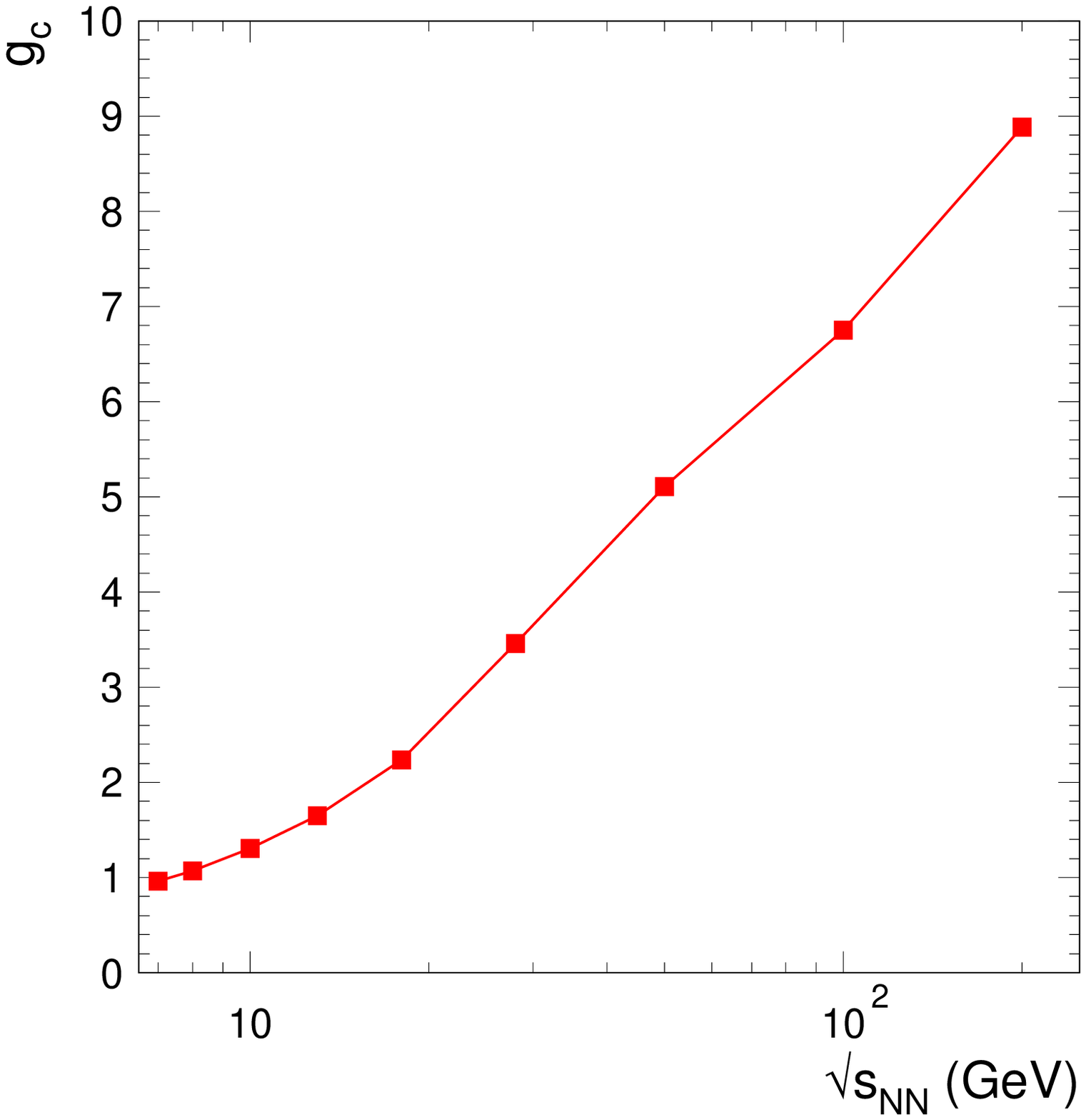}
\end{minipage} \end{tabular}
\caption{Energy dependence of the canonical suppression for charm, 
$I_1/I_0$ (left panel) and of the charm quark fugacity, $g_c$ (right 
panel).} 
\label{aa_fig1c}
\end{figure} 

For the studied energy range,  $\sqrt{s_{NN}}$=7-200 GeV, $T$ rises from 
151 to 161 MeV from $\sqrt{s_{NN}}$=7 to 12 GeV and stays constant for 
higher energies, while $\mu_b$ decreases from 434 to 22 MeV \cite{aat}.
The volume $V_{\Delta y=1}$ at midrapidity, shown in Fig.~\ref{aa_fig1b} 
\cite{aat} continuously rises from 760 to 2400 fm$^3$.  
Due to the strong energy dependence of charm production, Fig.~\ref{aa_fig1a},
the canonical suppression factor ($I_1/I_0$) varies from 1/30 to 1/1.2. 
Correspondingly, the charm fugacity $g_c$ increases from 0.96 to 8.9,
see Fig.~\ref{aa_fig1c}.

Before proceeding to discuss our results, we would like to emphasize some 
peculiar aspects of charm at low energies.
First, the assumption of charm equilibration can be questionable.  
In this exploratory study we have nevertheless assumed full thermalization.
At SPS and lower energies collision time, plasma formation time, and 
charmonium (or open charm hadrons) formation time are all of the same 
order \cite{bla}. 
Furthermore, the maximum plasma temperature may not exceed the $J/\psi$
dissociation temperature, $T_D$, although recent results \cite{moc} indicate 
that $T_D$ can be very close to $T_c$. 
Charmonia may be broken up by by gluons and by high energy nucleons still 
passing by from the collision. In this latter case cold nuclear suppression 
needs to be carefully considered (as discussed, e.g., in \cite{satz1,arleo}).
Consequently, our calculations, in which both charmonium formation before 
QGP production and cold nuclear suppression are neglected, may somewhat 
underestimate the charmonium production yield at SPS energies \cite{aa2}
and below.

We note that models that combine the 'melting scenario' with statistical 
hadronization have been proposed \cite{gra0}. Alternatively, charmonium 
formation by coalescence in the plasma \cite{the1,the2,gra,yan} as well 
as within transport model approaches \cite{zha,bra} has been considered.

\section{Energy dependence of charmed hadrons yield}

\begin{figure}[hbt]
\centering\includegraphics[width=1.02\textwidth,height=.58\textwidth]{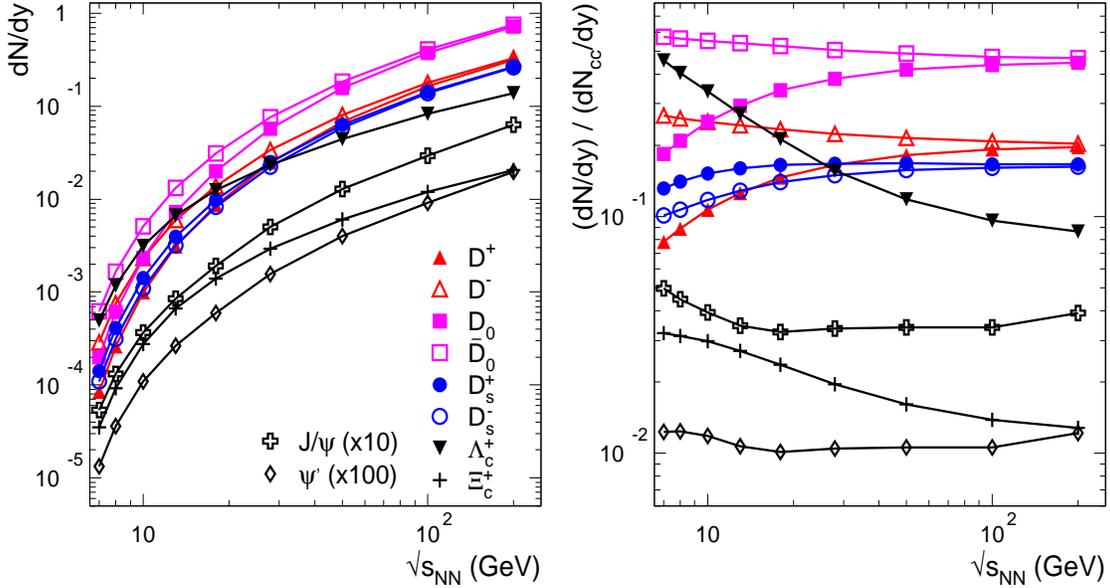}
\caption{Energy dependence of charmed hadron production at midrapidity.
Left panel: absolute yields, right panel: yields relative to the number
of $c\bar{c}$ pairs. Note, in both panels, the scale factors of 10 and 100 
for $J/\psi$ and $\psi'$ mesons, respectively.} 
\label{aa_fig1}
\end{figure}

Our main results are presented in Fig.~\ref{aa_fig1}.  The left panel shows
our predictions for the energy dependence of midrapidity yields for various
charmed hadrons.  Beyond the generally decreasing trend towards low energies
for all yields one notices first a striking behavior of the production of 
$\Lambda_c^+$ baryons: their yield exhibits a weaker energy dependence than 
observed for other charmed hadrons. In our approach this is caused by the 
increase in baryochemical potential towards lower energies (coupled with the 
charm neutrality condition). 
A similar behavior is seen for the $\Xi_c^+$ baryon.  
These results emphasize the importance of measuring, in addition to D-meson 
production, also the yield of charmed baryons to get a good measure of the 
total charm production cross section.
In detail, the production yields of D-mesons depend also on their quark 
content.

The differing energy dependences of the yields of charmed hadrons
are even more evident in the right panel of Fig.~\ref{aa_fig1}, where 
we show the predicted yields normalized to the number of initially produced
$c\bar{c}$ pairs.
Except very near threshold, the $J/\psi$  production yield per $c\bar{c}$ pair 
exhibits a slow increase with increasing energy. This increase is a consequence 
of the quadratic term in the J/$\psi$ yield equation discussed above. 
At LHC energy, the yield ratio J/$\psi/c\bar{c}$ approaches 
1\% \cite{aa2}, scaling linearly with $\sigma_{c\bar{c}}$ (for details see
\cite{aa4}). 
The $\psi'$ yield shows a similar energy dependence as the $J/\psi$,
except for our lowest energies, where the difference is due to the
decrease of temperature (see above).
We emphasize again that this model prediction, namely yields relative to 
$c\bar{c}$ pairs, is a robust result, as it is in the first order independent 
on the charm production cross section.
Due to the expected similar temperature, the relative abundance of open charm 
hadrons at LHC is predicted \cite{aa5} to be similar to that at RHIC energies.

\section{Effects of in-medium modification of charmed hadrons masses}

We consider two scenarios\footnote{The scenarios are constructed by
  modification of the constituent quark masses of light ($u$ and $d$) quarks in
  the charmed hadrons by fixed amounts. Reducing, for example, the light quark
  masses by 50 MeV will lower D-meson masses by 50 MeV and the $\Lambda_c
  (\Xi_c)$ mass by 100 (50) MeV.} for a possible mass change $\Delta m$ of open
charm hadrons containing light, $u$ or $d$, quarks: i) a common decrease of 50
MeV for all charmed mesons and their antiparticles and a decrease of 100 MeV
for the $\Lambda_c$ and $\Sigma_c$ baryons (50 MeV decrease for $\Xi_c$); ii)
a decrease of 100 MeV for all charmed mesons and a 50 MeV increase for their
antiparticles, with the same (scaled with the number of light quarks) scenario
as in i) for the baryons.  Scenario i) is more suited for an isospin-symmetric
fireball produced in high-energy collisions and was used in \cite{cas}, while
scenario ii) may be realized at low energies.  In both scenarios, the masses
of the $D_s$ mesons and of the charmonia are the vacuum masses. 
We also note that if one leaves all D-meson masses unchanged but 
allows their widths to increase, the resulting yields will increase by 11\%
(2.7\%) for a  width of 100 MeV (50 MeV).
If the in-medium widths exhibit tails towards low masses, as has been 
suggested by \cite{tol}, to first order the effect on thermal densities
is quantitatively comparable with that from a decrease in the pole mass.

\begin{figure}[htb]
\centering\includegraphics[width=1.03\textwidth]{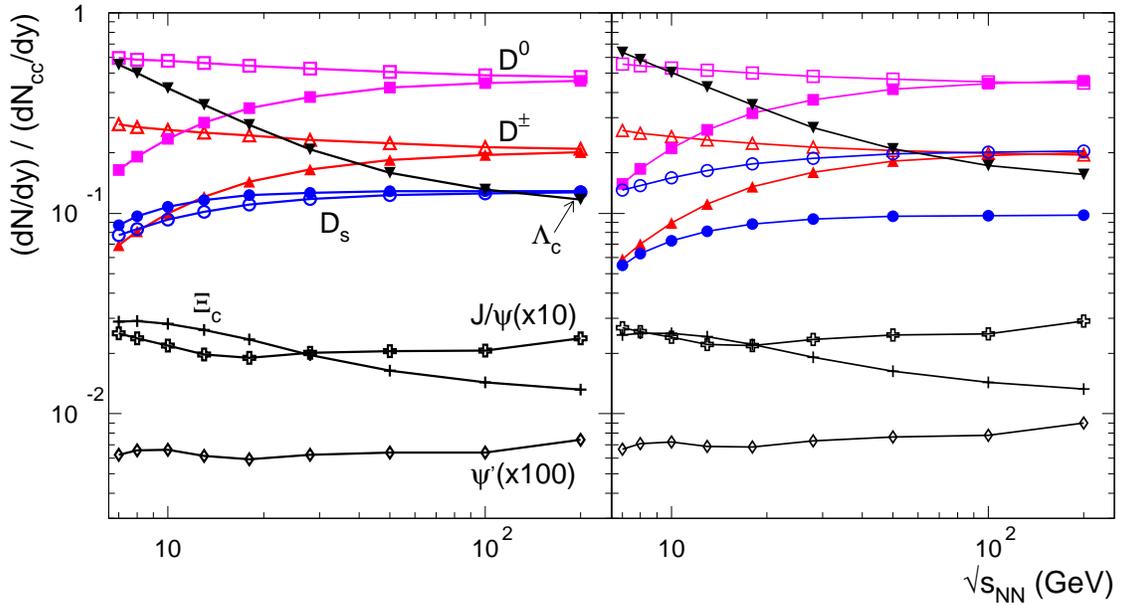}
\caption{Energy dependence of the yield of charmed hadrons relative to
the charm quark pair yield for two scenarios of the mass change 
(left panel for scenario i), right panel for scenario ii), see text).
For the D mesons, the full and open symbols are for particles and 
antiparticles, respectively. Note the factors 10 and 100
for the $J/\psi$ and $\psi'$ mesons, respectively.} 
\label{aa_fig2x}
\end{figure} 

The results for the two cases are presented in Fig.~\ref{aa_fig2x} as yields
relative to the number of initially-produced $c\bar{c}$ pairs.  As a result of
the redistribution of the charm quarks over the various species, the relative
yields of charmed hadrons may change. For example, in scenario i) the ratios 
of D-mesons are all close to those computed for vacuum masses
(Fig.~\ref{aa_fig1}), while for scenario ii) the changes in the relative 
abundances of the $D$ and $\bar{D}$ mesons are obvious.
In both cases the $\Lambda_c$/D ratio is increased.

As a result of the asymmetry in the mass shifts for particles and 
antiparticles assumed in scenario ii), coupled with the charm
neutrality condition, the production yields of $D^+_s$ and $D^-_s$ mesons 
are very different compared to vacuum masses.
Overall, however, charm conservation leads to rather small changes in the 
total yields.
We emphasize that, although the charm conservation equation is strictly 
correct only for the total cross section we expect within the framework 
of the statistical hadronization model, also little influence due to medium 
effects on distributions in rapidity and transverse momentum.
This is due to the fact that the crucial input into our model is 
$\ud N_{c\bar{c}}^{AuAu}/\ud y$ and there is no substantial D-meson 
rescattering after formation at the phase boundary. 

\begin{figure}[hbt]
\centering\includegraphics[width=.65\textwidth]{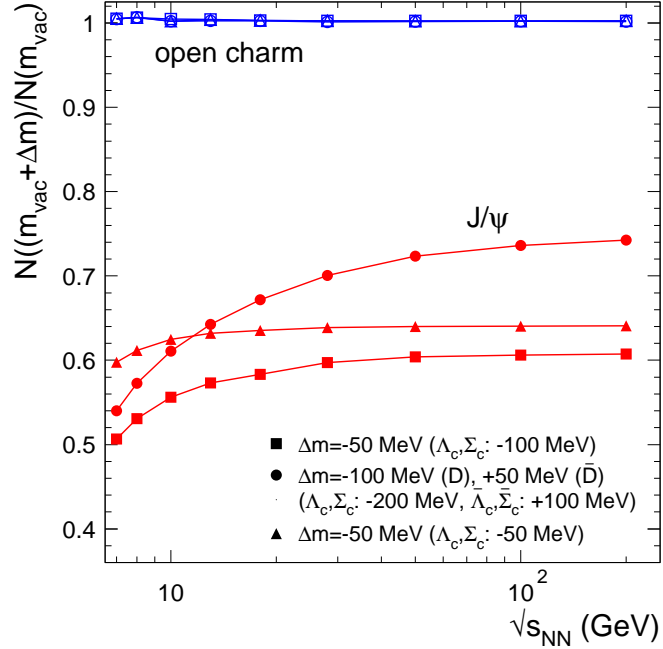}
\caption{Energy dependence of the relative change in the production
yield of open charm hadrons and of $J/\psi$ meson considering 
different scenarios for in-medium mass modifications (see text).} 
\label{aa_fig2}
\end{figure} 

In Fig.~\ref{aa_fig2} we demonstrate that the total open charm yield (sum over
all charmed hadrons) exhibits essentially no change if one considers mass
shifts, while the effect is large on charmonia. This is to be expected from
eq. \ref{aa:eq1}: as the masses of open charm mesons and baryons are reduced, 
the charm fugacity g$_c$ is changed accordingly to conserve charm.  
Consequently, since the open charm yields vary linearly with g$_c$, one 
expects little change with medium effects in this case. In contrast, the
yields of charmonia vary strongly, since they are proportional to $g_c^2$.
To demonstrate this we plot, in Fig.~\ref{aa_fig2}, the relative change 
of the yields with in-medium masses compared to the case of vacuum masses. 
For this comparison, we have added a third case, namely considering that 
the mass change of charmed baryons is the same as for the mesons.
Because of total charm conservation, with lowering of their masses the 
open charm hadrons eat away some of the charm quarks of the charmonia but, 
since the open charm hadrons are much more abundant, their own yield will 
hardly change. 

Note that the reduction of the J/$\psi$ yield in our model is quite different
from that assumed in \cite{zha,gra,hay,sib,fri}, where a reduction in
D-meson masses leads to the opening up of the decay of $\psi'$ and 
$\chi_c$ into $D\bar{D}$ and subsequently to a smaller J/$\psi$ yield
from feed-down from $\psi'$ and $\chi_c$.
In all the previous work the in-medium masses are considered in 
a hadronic stage, while our model is a pure QGP model, with in-medium mass 
modifications considered at the phase boundary.

\section{Conclusions}

We have investigated charmonium production in the statistical hadronization
model at lower energies. An interesting result is that the yield of charmed
baryons ($\Lambda_c$, $\Xi_c$) relative to the total $c\bar{c}$ yield
increases strongly with decreasing energy. Below $\sqrt{s_{NN}}$=10 GeV, the
relative yield of $\Lambda_c$ exceeds that of any D meson except $\bar{D}_0$, 
implying that an investigation of open charm production at low energies needs 
to include careful measurements of charmed baryons, a difficult experimental 
task.  
The charmonium/open charm yield rises only slowly from energies near threshold 
to reach $\sim$1\% at LHC energy.  Note that this ratio depends on the 
magnitude of the charm cross section, further underlining the importance 
to measure this quantity with precision.  We have also investigated the
effect of possible medium modifications of the masses of charmed hadrons.
Because of a separation of time scales for charm quark and charmed hadron
production, the overall charmed meson and baryon cross section is very little 
affected by in-medium mass changes, if charm conservation is taken into account.
Measurable effects are predicted for the yields of charmonia. 
These effects are visible at all beam energies and are more pronounced 
towards threshold.

\section*{Acknowledgments}
K.R. acknowledges partial support from the Polish Ministry of Science (MEN).


\begin{thebibliography}{99}

\bibitem{satz} T. Matsui, H. Satz, Phys. Lett. B 178 (1986) 416.

\bibitem{satz2} F. Karsch, D. Kharzeev, H. Satz, Phys. Lett. B 637 (2006) 75
[hep-ph/0512239]; H. Satz, Nucl. Phys. A 783 (2007) 249 [hep-ph/0609197].

\bibitem{aa1} A. Andronic, P. Braun-Munzinger, K. Redlich, J. Stachel, 
Phys. Lett. B 571 (2003) 36 [nucl-th/0303036].

\bibitem{aa2} A. Andronic, P. Braun-Munzinger, K. Redlich, J. Stachel, 
Nucl. Phys. A 789 (2007) 334 [nucl-th/0611023].

\bibitem{aa3} A. Andronic, P. Braun-Munzinger, K. Redlich, J. Stachel, 
Phys. Lett. B 652 (2007) 259 [nucl-th/0701079].

\bibitem{pbm1} P. Braun-Munzinger, J. Stachel,
Phys. Lett. B 490 (2000) 196 [nucl-th/0007059];
Nucl. Phys. A 690 (2001) 119c [nucl-th/0012064].

\bibitem{phe1} A. Adare et al. (PHENIX),  Phys. Rev. Lett. 98 (2007) 232301 
[nucl-ex/0611020].

\bibitem{aa5} A. Andronic, P. Braun-Munzinger, K. Redlich, J. Stachel, 
arXiv:0707.4075 

\bibitem{cbm1} P. Senger, J. Phys. Conf. Series 50 (2006) 357.
\bibitem{tol} L. Tolos,  J. Schaffner-Bielich, H. St\"ocker,
Phys. Lett. B 635 (2006) 85 [nucl-th/0509054].

\bibitem{tsu} K. Tsushima, D.H. Lu, A.W. Thomas, K. Saito, R.H. Landau,
Phys. Rev. C 59 (1999) 2824 [nucl-th/9810016];
\bibitem{sib1} A. Sibirtsev, K. Tsushima, A.W. Thomas, 
Eur. Phys. J. A 6 (1999) 351 [nucl-th/9904016].
\bibitem{sib} A. Sibirtsev, K. Tsushima, K. Saito, A.W. Thomas,
Phys. Lett. B 484 (2000) 23 [nucl-th/9904015].
\bibitem{hay} A. Hayashigaki, Phys. Lett. B 487 (2000) 96 
[nucl-th/0001051].
\bibitem{cas} W. Cassing, E.L. Bratkovskaya, A. Sibirtsev, 
Nucl. Phys. A 691 (2001) 753 [nucl-th/0010071].
\bibitem{fri} B. Friman, S.H. Lee, T. Song, Phys. Lett. B 548 (2002) 153
[nucl-th/0207006].
\bibitem{lutz} M.F.M. Lutz, C.L. Korpa, Phys. Lett. B 633 (2006) 43
[nucl-th/0510006].
\bibitem{mor} K. Morita, S.H. Lee, arXiv:0704.2021.

\bibitem{aa4} A. Andronic, P. Braun-Munzinger, K. Redlich, J. Stachel, 
arXiv:0708.1488 [nucl-th].


\bibitem{aat} A. Andronic, P. Braun-Munzinger, J. Stachel, 
Nucl. Phys. A 772 (2006) 167 [nucl-th/0511071].

\bibitem{bro} M. Michalec, W. Florkowski, W. Broniowski, 
Phys. Lett. B 520 (2001) 213 [nucl-th/0103029].
\bibitem{zschiesche} D. Zschiesche, S. Schramm, J. Schaffner-Bielich,
  H. St\"ocker, W. Greiner, Phys. Lett. B 547 (2002) 7.

\bibitem{gor} M.I. Gorenstein, A.P. Kostyuk, H. St\"ocker, W. Greiner,
Phys. Lett. B 509 (2001) 277 [hep-ph/0010148].

\bibitem{rv1} R. Vogt, Int. J. Mod. Phys. E 12 (2003) 211 [hep-ph/0111271].

\bibitem{lou} C. Louren\c co, H. W\"ohri, Phys. Rept. 433 (2006) 127
[hep-ph/0609101].

\bibitem{phe3} A. Adare et al. (PHENIX), Phys. Rev. Lett. 97 (2006) 252002 
[hep-ex/0609010].

\bibitem{herab} I. Abt et al. (HERA-B), Phys. Lett. B 638 (2006) 407
[hep-ex/0512029].

\bibitem{cac} M. Cacciari, P. Nason, R. Vogt, Phys. Rev. Lett. 95 (2005) 
122001 [hep-ph/0502203]. 

\bibitem{pdg} W.-M. Yao et al., J. Phys. G 33 (2006) 1
[{\tt http://pdg.lbl.gov/}].

\bibitem{pbm2} P. Braun-Munzinger, D. Mi\'skowiec, A. Drees, C. Louren\c co,
Eur. Phys. J. C 1 (1998) 123 [nucl-ex/9704011].

\bibitem{bla} F. Karsch and R. Petronzio, Phys. Lett. B 193 (1987) 105;
J.-P. Blaizot and J.-Y. Ollitrault, Phys. Rev. D 39 (1989) 232.

\bibitem{moc} A. Mocsy, P. Petrecky, arXiv:0705.2559, arXiv:0706.2183.

\bibitem{satz1} H. Satz, J. Phys. G 32 (2006) R25 [hep-ph/0512217].
\bibitem{arleo} F. Arleo, V. Tram, hep-ph/0612043.

\bibitem{gra0} L. Grandchamp, R. Rapp, Phys. Lett. B 523 (2001) 60
[hep-ph/0103124]; Nucl. Phys. A 709 (2002) 415 [hep-ph/0205305].
\bibitem{the1} R.L. Thews, M. Schroedter, J. Rafelski, Phys. Rev. C63 (2001)
  054905. 
\bibitem{the2} R.L. Thews, M.L. Mangano, Phys. Rev. C 73 (2006) 014904
[hep-ph/0505055]. 
\bibitem{gra} L. Grandchamp, R. Rapp, G.E. Brown, Phys. Rev. Lett. 92 (2004) 
212301 [hep-ph/0306077].
\bibitem{yan} L. Yan, P. Zhuang, N. Xu, Phys. Lett. B 607 (2005) 107
[nucl-th/0411093]; Phys. Rev. Lett. 97 (2006) 232301 [nucl-th/0608010].

\bibitem{zha} B. Zhang, C.M. Ko, B.A. Li, Z.W. Lin, S. Pal, 
Phys. Rev. C 65 (2002) 054909 [nucl-th/0201038].
\bibitem{bra} E.L. Bratkovskaya, W. Cassing, H. St\"ocker, 
Phys. Rev. C 67 (2003) 054905 [nucl-th/0301083];
O. Linnyk, E. L. Bratkovskaya, W. Cassing, H. St\"ocker,
Nucl. Phys. A 786 (2007) 183 [nucl-th/0612049].


\end{thebibliography}
\end{document}